\documentclass[a4paper,14pt]{article}
\usepackage[english]{babel}
\pdfoutput=1 

\usepackage{jheppub} 

\usepackage[T1]{fontenc} 

\usepackage[T1]{fontenc}
\usepackage[utf8]{inputenc}

\usepackage{mathrsfs}

\usepackage{booktabs}

\usepackage{comment}

\usepackage{cases}

\usepackage{mathtools}

\newcommand{\bi}{\begin{itemize}}
    \newcommand{\ei}{\end{itemize}}
\newcommand{\bea}{\begin{eqnarray}}
    \newcommand{\eea}{\end{eqnarray}}
\newcommand{\bt}{\begin{tabular}}
    \newcommand{\et}{\end{tabular}}
\newcommand{\bc}{\begin{center}}
    \newcommand{\ec}{\end{center}}

\newcommand{\be}{\begin{equation}}
    \newcommand{\ee}{\end{equation}}
\newcommand{\ba}{\begin{array}}
    \newcommand{\ea}{\end{array}}

\def\bbox{{\,\lower0.9pt\vbox{\hrule \hbox{\vrule height 0.2 cm
                \hskip 0.2 cm \vrule height 0.2 cm}\hrule}\,}}
\newcommand{\dsl}{\pa \kern-0.5em /}

\makeatletter \@addtoreset{equation}{section} \makeatother

\def\slashchar#1{\setbox0=\hbox{$#1$}           
    \dimen0=\wd0                                 
    \setbox1=\hbox{/} \dimen1=\wd1               
    \ifdim\dimen0>\dimen1                        
    \rlap{\hbox to \dimen0{\hfil/\hfil}}      
    #1                                        
    \else                                        
    \rlap{\hbox to \dimen1{\hfil$#1$\hfil}}   
    /                                         
    \fi}

\pdfoutput=1

\title{\boldmath  On Born-Infeld-type higher-spin interactions and supersymmetry}

\author[]{Nikita~Zaigraev}

\affiliation[]{Bogoliubov Laboratory of Theoretical Physics, JINR,\\
	Joliot-Curie st. 6, 141980 Dubna, Moscow region, Russia}
\affiliation[]{Moscow Institute of Physics and Technology,\\ Institutskiy per. 9, 141700 Dolgoprudny, Moscow region, Russia}

\emailAdd{nikita.zaigraev@phystech.edu}

\abstract{
	We develop a simple Lorentz-covariant construction of the
	Born--Infeld-type cubic interactions $(s, s_1, s_2)$ of massless 
	higher-spin fields with $s + s_1 + s_2$ derivatives in four-dimensional 
	flat space. For arbitrary spins $s \ge s_1 \ge s_2$, the vertex is 
	formulated in terms of a conserved higher-spin current and its trace. 
	In contrast to the famous Berends--Burgers--van Dam currents, these 
	currents are constructed from higher-spin Weyl tensors of a single chirality, 
	schematically $J \sim \partial C \partial C$. Starting with a general 
	ansatz, we derive recurrence relations fixing the current coefficients, 
	identify the unique non-trivial conserved current after eliminating 
	trivial (fake) vertices, and show that in the range $s \le s_1 + s_2$ 
	the interaction admits a manifestly gauge-invariant representation in 
	terms of higher-spin Weyl tensors. The construction is then extended 
	to $\mathcal{N}=1$ and $\mathcal{N}=2$ supersymmetry using the 
	corresponding off-shell superfield formalisms. In the $\mathcal{N}=1$ 
	case, we obtain the general conserved integer superspin $s$ 
	supercurrent with manifest supersymmetry, which is bilinear in the Weyl 
	supertensors of superspins $Y_1$ and $Y_2$.
	For identical superfields, $Y_1 = Y_2$, this supercurrent 
	vanishes for odd $s$ and is in agreement with earlier results by Gates and 
	Koutrolikos. In the $\mathcal{N}=2$ case, using harmonic superspace, 
	we construct the principal higher-spin supercurrent whose coefficients 
	are uniquely fixed by the conservation equations. In contrast to the 
	$\mathcal{N}=0,1$ cases, for identical $\mathcal{N}=2$ superfields, 
	the corresponding vertex does not vanish. The results provide a 
	systematic framework for Born--Infeld-type higher-spin interactions 
	and their supersymmetric extensions.
}

\makeatletter
\gdef\@fpheader{}
\makeatother

\begin{document}

\maketitle
\flushbottom


\section{Introduction}

The construction of cubic higher-spin interactions is the first important step toward studying the structure of consistent interactions.
Consistent cubic higher-spin interactions in flat $4D$ spacetime were classified in the works of Metsaev \cite{Metsaev:2005ar, Metsaev:2007rn, Metsaev:2022yvb}. According to this classification, there exist two types of parity-invariant vertices $(s, s_1, s_2)$\footnote{We assume that $s\geq s_1\geq s_2$.}, which can be represented in manifestly Lorentz covariant form: either with $s + s_1 - s_2$ or  $s + s_1 + s_2$
derivatives.
Such interactions appear in the Vasiliev theory \cite{Vasiliev:1990en, Vasiliev:1992av}, as was shown in papers \cite{Gelfond:2017wrh, Misuna:2017bjb, Tatarenko:2024csa}.
 Vertices with the maximum number of derivatives are known in the literature as the Born–Infeld (BI) or non-minimal interactions \cite{Bekaert:2010hw, Ponomarev:2022vjb, Pekar:2023nev}
 and are necessary for constructing consistent interactions at higher orders \cite{Metsaev:1991mt}.

\smallskip

Interactions of higher-spin fields are being actively investigated; see, e.g., \cite{Didenko:2023vna, Vasiliev:2025hfh, Didenko:2024zpd, Didenko:2026nag}.
As for the Lorentz-covariant realization of the Born–Infeld-type vertices, to the best of our knowledge, they were first discussed in the work \cite{Damour:1987fp}.
Further discussion of such vertices can be found in papers  \cite{Boulanger:2008tg, Manvelyan:2010jr}. Also, the construction of such vertices was considered within the BRST approach~\cite{Buchbinder:2022yiz}, but the explicit field structure of such an interaction was found only for the simplest vertex with two scalar fields.

\smallskip

In this paper, we present a simple, explicitly Lorentz-covariant construction of the Born–Infeld-type vertices in 4D for arbitrary spin values $s \geq s_1 \geq s_2$ and then generalize it to the cases of 
$\mathcal{N}=1$ and 
$\mathcal{N}=2$ supersymmetry using the corresponding superfield formulations.

\smallskip

Let us begin with a schematic description of the BI vertices.
We will describe the higher-spin fields in the language of two-component spinors. For integer  spin $s$, the
 theory is described by two unconstrained fields defined up to the gauge freedom\footnote{In this formula and in what follows, we assume full symmetrization over all dotted and undotted spinor indices.}:
 \begin{equation}\label{eq: HS transf}
 	\delta \Phi_{\alpha(s)\dot{\alpha}(s)} = \partial_{\alpha\dot{\alpha}} \xi_{\alpha(s-1)\dot{\alpha}(s-1)},
 	\qquad
 		\delta \Phi_{\alpha(s-2)\dot{\alpha}(s-2)} = \partial^{\beta\dot{\beta}} \xi_{(\beta\alpha(s-1))(\dot{\beta}\dot{\alpha}(s-1))}.
 \end{equation}	
 In terms of these fields, one can construct  two gauge-invariant tensors — self-dual and anti-self-dual higher-spin  Weyl tensors, which remain non-zero when the free equations of motion are imposed:
 \begin{equation}\label{eq: Weyl curv}
 	C_{\alpha(2s)} := (\partial^s)_{\alpha(s)}^{\dot{\beta}(s)}  \Phi_{\alpha(s)\dot{\beta}(s)},
 	\qquad
 	\bar{C}_{\dot{\alpha}(2s)}
 	:=
 	(\partial^s)_{\dot{\alpha}(s)}^{\beta(s)}  \Phi_{\beta(s)\dot{\alpha}(s)}.
 \end{equation}	
 Higher-spin tensors contain $s$ space-time derivatives and being the main building blocks for Born–Infeld vertices.  Analogously, Weyl tensors are constructed for arbitrary half-integer spin as well; see, for example \cite{Zaigraev:2026jvu}. On the mass shell, higher-spin Weyl tensors satisfy\footnote{Here and below 
 	$\approx$ denotes equality modulo the free equations of motion.}
 \begin{equation}\label{eq: weyl cons} \partial^\beta_{\dot{\beta}} C_{\beta\alpha(2s-1)} \approx 0,
 	\qquad 
 	\partial^{\dot{\beta}}_\alpha \bar{C}_{\dot{\beta}\dot{\alpha}(2s-1)} \approx 0.
 \end{equation}

Using Weyl tensors, it is straightforward to show that higher-spin BI vertices can be divided into two subtypes, containing either three explicit Weyl tensors or two Weyl tensors inside a conserved current,  depending on relations between spins.

\smallskip

\textbf{1.} In the case  $s \leq s_1 + s_2$  BI interactions can be constructed in terms of higher-spin Weyl tensors \eqref{eq: Weyl curv}.
  Let us denote $k = s_1+s_2-s $, then
\begin{equation}\label{eq: C3}
\mathcal{L}_{(s,s_1,s_2)} \sim	C_s^{\alpha(2s)} (C_{s_1})_{\alpha(2s_1-k)\beta(k)} (C_{s_2})^{\beta(k)}_{\alpha(2s_2-k)}.
\end{equation}

\textbf{2.}  It is straightforward to see that in the case $s> s_1 + s_2$ it is impossible to construct vertices solely in terms of higher-spin Weyl tensors. Vertices of this type can be constructed using conserved higher-spin currents:
\begin{subequations}
\begin{equation}
 \mathcal{L}_{(s,s_1,s_2)} \sim	\Phi^{\alpha(s)\dot{\alpha}(s)} J_{\alpha(s)\dot{\alpha}(s)}
	+
	  \Phi^{\alpha(s-2)\dot{\alpha}(s-2)} T_{\alpha(s-2)\dot{\alpha}(s-2)},
\end{equation}	
which, on the mass shell of the spin-$s_1$
and spin-$s_2$
fields, must satisfy the conservation equation
\begin{equation}\label{eq: cons equation}
	\partial^{\beta\dot{\beta}}
	J_{(\alpha(s-1)\beta) (\dot{\alpha}(s-1)\dot{\beta})} + \partial_{\alpha\dot{\alpha}} T_{\alpha(s-2)\dot{\alpha}(s-2)} \approx 0.
\end{equation}	
\end{subequations}
It is important that these currents must be constructed from Weyl tensors of the same  chirality $J \sim \partial C \partial C$, which distinguishes them from the Berends-Burgers-van Dam higher-spin currents \cite{Berends:1985xx} and their generalizations \cite{Gelfond:2006be, Manvelyan:2009vy, Zinoviev:2010cr, Zaigraev:2026xzr}, which are constructed from Weyl tensors of different chirality $J\sim \partial C \partial \bar{C}$.
To the best of our knowledge, such currents have not been studied in the literature,
although in the special case of equal spins they are contained in the components of the  $\mathcal{N}=1$ supercurrents constructed by Gates and Koutrolikos \cite{Gates:2019cnl}.

\medskip

If we consider theories with supersymmetry in flat $4D$ space, the cubic interactions were also classified by Metsaev for minimal $\mathcal{N}=1, 4D$ supersymmetry \cite{Metsaev:2019dqt}\footnote{There also exists a generalization of this classification to extended 
	$\mathcal{N}=4\mathbb{N}$ supersymmetries \cite{Metsaev:2019aig}.}. According to this classification, there exist several types of cubic $\mathcal{N}=1$ vertices.
Let us first introduce the following notation:
\begin{equation}
	\begin{split}
		& \Psi_{s} \quad
		(s+\frac{1}{2}, s)
		\quad\qquad
		  \text{integer superspin $s$ multiplet}
		\\
		& \Phi_s 
		\quad
		(s, s-\frac{1}{2})
		 \quad\quad\quad
		   \text{half-integer superspin $s-\frac{1}{2}$ multiplet}
	\end{split}	
\end{equation}	
Then it is convenient to present the classification in a tabular form. In the table  \ref{tab:metsaev-n1}, we present the Metsaev classification and have also marked the vertices that are guaranteed to admit a Lorentz-covariant formulation. In particular, we highlight that there are two types  of $\mathcal{N}=1$ supersymmetric Born–Infeld-type vertices (2a and 4a). A distinctive feature of these vertices is that they necessarily involve a supermultiplet of integer superspin.


\begin{table}[htbp]
	\begin{center}
		\caption{Classification of cubic interaction vertices for \(\mathcal{N} = 1\) massless supermultiplets of arbitrary spin
			in \(4D\) flat space (Metsaev \cite{Metsaev:2019dqt}).  \(K^{\text{cov}}_{BBB}\) and \(K^{\text{cov}}_{FFB}\) are the numbers of derivatives in the
			covariant formulation for a vertex with three bosonic fields and with two fermionic plus one bosonic field,
			respectively.}
		\label{tab:metsaev-n1}
		\medskip
		\makebox[\textwidth][c]{%
			\begin{tabular}{@{}cccccc@{}}
				\toprule
				Vertex type &  & Allowed spins & \(K^{\text{cov}}_{BBB}\) & \(K^{\text{cov}}_{FFB}\) & Lorentz covariant form \\
				\midrule
				\multicolumn{5}{@{}l}{1.  $\Phi\Phi\Phi $} \\
				$\Phi_{s_1}\Phi_{s_2}\Phi_{s_3} $
				& 
				& \(s_1\ge 1,\ s_2\ge 1,\ s_3\ge 1\)
				& \(s_1+s_2-s_3\)
				& \(s_1+s_2-s_3-1\) &$\checkmark$ \\
				& & \(s_1+s_2-s_3\ge 1\) &  &  \\[6pt]
					\midrule
				\multicolumn{5}{@{}l}{2. $\Phi\Phi\Psi$ }\\[2pt]
				2a. $\Phi_{s_1}\Phi_{s_2}\Psi_{s_3}$
				& 
				& \(s_1\ge 1,\ s_2\ge 1,\ s_3\ge 0\)
				& \(s_1+s_2+s_3\)
				& \(s_1+s_2+s_3-1\) &$\checkmark$  \\
				& & \(s_1+s_2+s_3\ge 1\) & & \\[3pt]
				2b. $\Phi_{s_1}\Phi_{s_2}\Psi_{s_3}$
				& 
				& \(s_1\ge 1,\ s_2\ge 1,\ s_3\ge 0\)
				& \(s_1-s_2-s_3\)
				& \(s_1-s_2-s_3-1\) \\
				& & \(s_1-s_2-s_3\ge 1\) & & \\[3pt]
				2c. $\Psi_{s_1}\Phi_{s_2}\Phi_{s_3}$
				& 
				& \(s_1\ge 0,\ s_2\ge 1,\ s_3\ge 1\)
				& ---
				& \(s_1-s_2-s_3\) \\
				& & \(s_1-s_2-s_3\ge 0\) & & \\[6pt]
					\midrule
				\multicolumn{5}{@{}l}{3.  $\Phi\Psi\Psi$} \\[2pt]
				3a. $\Phi_{s_1}\Psi_{s_2}\Psi_{s_3}$
				& 
				& \(s_1\ge 1,\ s_2\ge 0,\ s_3\ge 0\)
				& \(s_1+s_2-s_3\)
				& \(s_1+s_2-s_3-1\) &$\checkmark$ \\
				& & \(s_1+s_2-s_3\ge 1\) & & \\[3pt]
				3b. $\Psi_{s_1}\Psi_{s_2}\Phi_{s_3}$
				& 
				& \(s_1\ge 0,\ s_2\ge 0,\ s_3\ge 1\)
				& ---
				& \(s_1+s_2-s_3\) &$\checkmark$ \\
				& & \(s_1+s_2-s_3\ge 0\) & & &\\[6pt]
				
					\midrule
					
				\multicolumn{5}{@{}l}{4. $\Psi\Psi\Psi$} \\[2pt]
				4a. $\Psi_{s_1}\Psi_{s_2}\Psi_{s_3}$
				& 
				& \(s_1\ge 0,\ s_2\ge 0,\ s_3\ge 0\)
				& ---
				& \(s_1+s_2+s_3\) &$\checkmark$  \\
				& & \(s_1+s_2+s_3\ge 0\) & & \\[3pt]
				4b. $\Psi_{s_1}\Psi_{s_2}\Psi_{s_3}$
				& 
				& \(s_1\ge 0,\ s_2\ge 0,\ s_3\ge 0\)
				& \(s_1-s_2-s_3\)
				& \(s_1-s_2-s_3-1\) \\
				& & \(s_1-s_2-s_3\ge 1\) & & \\
				\bottomrule
			\end{tabular}%
		}
	\end{center}
\end{table}

\medskip

While it is tempting to assume that cubic vertices from non-supersymmetric higher-spin theories can be straightforwardly supersymmetrized, this holds only on-shell, where supersymmetry invariance must be checked and closes only with the equations of motion. Demanding off-shell (manifest) supersymmetry forces the use of the entire supermultiplet, including auxiliary fields that balance bosonic and fermionic degrees of freedom and ensure the supersymmetry algebra closes without equations of motion. This structure is automatically realized in superspace \cite{BK, Gates:1983nr, Wess:1992cp, 18}. 

\smallskip

The superfield formulation of 
$\mathcal{N}=1, 4D$ higher-spin theories was constructed in works \cite{Kuzenko:1993jp, Kuzenko:1993jq} (see also \cite{BK} for a pedagogical review). On the basis of this formalism, various consistent $\mathcal{N}=1$ cubic vertices with manifest supersymmetry have been constructed for the interaction of higher-spin fields with matter \cite{Kuzenko:2017ujh, Hutomo:2017phh, Koutrolikos:2017qkx, Buchbinder:2018wwg, Buchbinder:2018gle} (and generalized to AdS background in \cite{Buchbinder:2018nkp}), a special class of abelian vertices for interactions between higher spins \cite{Buchbinder:2018wzq}, and also a special class of the Born–Infeld-type interactions \cite{Gates:2019cnl}. The construction of vertices in these works was based on the construction of gauge-invariant $\mathcal{N}=1$ supercurrents.
We also note that non-Abelian supersymmetric vertices ($s_1<s_2+s_3$) were constructed in the work \cite{Khabarov:2020deh} within the component approach in the framework of the Fradkin–Vasiliev formalism \cite{Fradkin:1986qy, Fradkin:1987ks, Vasiliev:2011knf, Khabarov:2020bgr}.

\smallskip

The  $\mathcal{N}=2, 4D$ superfield formalism for describing higher spins in harmonic superspace was constructed in the work \cite{Buchbinder:2021ite} (for review see \cite{Buchbinder:2026nwu, Buchbinder:2025ceg}). Harmonic superspace \cite{18, Galperin:1984av} permits to give a geometric interpretation of higher-spin superfields as vielbeins in the covariantized harmonic derivative, which made it possible to construct cubic interaction vertices with the hypermultiplet \cite{Buchbinder:2022kzl, Buchbinder:2022vra, Buchbinder:2024pjm}. In a recent work \cite{Buchbinder:2025yef}, it was shown that these results admit a generalization to $\mathcal{N}=1, 5D$ higher-spin theories.
A discussion of the corresponding supercurrents in the frameworks of conformal and projective superspaces can be found in 
\cite{Kuzenko:2023vgf, Kuzenko:2024vms}.
Within the harmonic  framework, the most general Abelian $\mathcal{N}=2$ cubic higher-spin interactions with manifest 
$\mathcal{N}=2$ supersymmetry off-shell were constructed in papers \cite{Zaigraev:2024ryg, Zaigraev:2026jvu, Zaigraev:2026xzr, Zaigraev:2025xqo}.

\smallskip

Based on the known 
$\mathcal{N}=1$ and 
$\mathcal{N}=2$ superspace formalisms for higher-spin supermultiplets, we will construct the most general Born–Infeld-type vertices with the manifest supersymmetric invariance.

\smallskip

This paper is organized as follows.
In Section \ref{sec: 2}, we analyze a simple example of the $(3,1,1)$ vertex, which illustrates the main subtleties that arise in the general case and in supersymmetric generalizations. In Section \ref{sec: general current} we present a conserved current corresponding to an arbitrary Born–Infeld-type vertex and show that, when condition $s< s_1+s_2$ is satisfied, this interaction can be written in a manifestly gauge-invariant form in terms of higher-spin Weyl tensors.
 Section \ref{sec: N=1 BI} is devoted to the generalization of these results to the case of supersymmetry.
In Section \ref{sec: N=2 gen}, we analyse $\mathcal{N}=2$ Born-Infeld interactions.
In the concluding Section \ref{sec: concl}, we summarize the results obtained.


\section{$(3,1,1)$  vertex with five derivatives}\label{sec: 2}

The purpose of this section is to illustrate the main features of the
general construction on the simplest non-trivial example, the
$(3,1,1)$ vertex with five derivatives. We  consider three cases:
$J\sim CC$ (two identical spin-1 fields), $J\sim C\bar C$
(different chiralities), and $J\sim CB$ (two distinct spin-1 fields).
The first two cases yield only fake vertices, while the third gives
a non-trivial Born--Infeld vertex.

\smallskip

\noindent \textbf{A.  $J \sim C C$}

\smallskip

We begin with the example in which the interaction involves two identical spin-1 fields.
Here and in what follows, we will use the notation for the symmetrized derivatives of the spin-$s$ Weyl tensor
\begin{equation}
	 \left( \partial^p C_s \right)_{\alpha(2s+p) \dot{\alpha}(p)}
	 : = 
	 \left( \partial^p\right)_{(\alpha(p)\dot{\alpha}(p)} \left( C_s \right)_{\alpha(2s))}. 
\end{equation}	
If there is no symmetrization over at least one index, the expression vanishes on the mass shell.

\smallskip

Then the general ansatz with 5 derivatives for the traceless spin-3 current constructed in terms of one spin-1 field is given by
\begin{subequations}
\begin{equation}\label{eq: CC with one spin 1}
	\mathbf{J}^{CC}_{\alpha(3)\dot{\alpha}(3)}
	=
	a_1 \left(C_1 \right)_{\beta(2)} \left( \partial^3 C_1 \right)^{\beta(2)}_{\alpha(3)\dot{\alpha}(3)}
	+
	a_2
	\left( \partial C_1 \right)_{\alpha\beta(2)\dot{\alpha}} (\partial^2 C_1 )^{\beta(2)}_{\alpha(2)\dot{\alpha}(2)},
\end{equation}	
where $a_1$ and $a_2$ are arbitrary complex coefficients.
The divergence of this current on spin-$1$ shell $(\partial_{\dot{\alpha}}^\alpha (C_1)_{\alpha(2)} \approx 0)$ takes the form
\begin{equation}
	\partial^{\rho\dot{\rho}} \mathbf{J}^{CC}_{\rho\alpha(2)\dot{\rho} \dot{\alpha}(2)}
	\approx
	\left[a_1 + \frac{1}{9}a_2\right]
		\left(\partial C_1  \right)_{\beta(3)\dot{\beta}} \left( \partial^3 C_1 \right)^{\beta(3)\dot{\beta}}_{\alpha(2)\dot{\alpha}(2)}
		+
	\frac{4}{9}a_2
	\left( \partial^2 C_1 \right)^{\rho\dot{\rho}}_{\alpha\beta(2)\dot{\alpha}} (\partial^2 C_1 )^{\beta(2)}_{\alpha\rho\dot{\alpha}\dot{\rho}}.
\end{equation}	
\end{subequations}
As we see, it is impossible to achieve the vanishing of this divergence.

Therefore, we consider a general ansatz for the current trace:
\begin{subequations}
\begin{equation}
	\mathbf{T}^{CC}_{\alpha\dot{\alpha}}
	=
 b	(\partial C_1)_{\beta(3)\dot{\beta}} (\partial^2 C_1)^{\beta(3)\dot{\beta}}_{\alpha\dot{\alpha}},
\end{equation}	
which gives a correction to the conservation law
\begin{equation}
	\partial_{\alpha\dot{\alpha}} \mathbf{T}^{CC}_{\alpha\dot{\alpha}}
	=
	 b	(\partial C_1)_{\beta(3)\dot{\beta}} (\partial^3 C_1)^{\beta(3)\dot{\beta}}_{\alpha(2)\dot{\alpha}(2)}
		+
 b	 	(\partial^2 C_1)_{\alpha\beta(3)\dot{\alpha}\dot{\beta}} (\partial^2 C_1)^{\beta(3)\dot{\beta}}_{\alpha\dot{\alpha}}.
\end{equation}	
\end{subequations}
	
From this, we find that the spin-3 current is conserved if one fixes  
\begin{equation}
	\begin{cases}
		a_1 + \frac{1}{9} a_2 + b = 0,
		\\
		\frac{4}{9}a_2 + b = 0
	\end{cases}	
	\quad
	\Rightarrow
	\quad
	a_1 = - \frac{3}{4} b,\;\;
	\text{and}\;\;
	a_2 = - \frac{9}{4} b.
\end{equation}	
Thus, we conclude that in this example there is one complex conserved current.
It is straightforward to see that the vertex obtained in this way cannot be written solely in terms of curvatures. Indeed, such a vertex depends on a spin-$3$ field $\Phi_{\alpha\dot{\alpha}}$ that does not enter the spin-3 Weyl tensor \eqref{eq: Weyl curv}.

\smallskip

Let us now consider a gauge-invariant \textit{fake vertex} containing curvatures, constructed from spin-3 fields and vanishing on the spin-$3$ mass shell:
\begin{equation}\label{eq: fake 5-1}
	\mathcal{L}^{CC}_{\text{fake}}
	\sim	\mathcal{R}^{\alpha\dot{\alpha}}_{(s=3)}
	(C_1)_{\beta(2)} (\partial C_1)^{\beta(2)}_{\alpha\dot{\alpha}},
\end{equation}	
Here the gauge-invariant Fronsdal-de Wit-Freedman tensors $\mathcal{R}$ for an arbitrary spin $s$ are defined by
\begin{subequations}\label{eq: R curv}
	\begin{equation}\label{eq: curv 1}
		\begin{split}
			\mathcal{R}_{\alpha(s)\dot{\alpha}(s)}
			:=\;&
			\Box \Phi_{\alpha(s)\dot{\alpha}(s)} -2s\, \partial_{\alpha\dot{\alpha}} \partial^{\beta\dot{\beta}}  \Phi_{(\alpha(s-1)\beta)(\dot{\alpha}(s-1)\dot{\beta})} 
			\\&
			+
			\frac{2(s-1)^2}{s} \partial_{\alpha\dot{\alpha}}  \partial_{\alpha\dot{\alpha}} \Phi_{\alpha(s-2)\dot{\alpha}(s-2)},
		\end{split}
	\end{equation}	
	\begin{equation}\label{eq: curv 2}
		\begin{split}
			\mathcal{R}_{\alpha(s-2)\dot{\alpha}(s-2)}
			:=\;&
			\Box \Phi_{\alpha(s-2)\dot{\alpha}(s-2)}
			+
			\frac{2(s-2)^2}{2s-1} \partial_{\alpha\dot{\alpha}} \partial^{\beta\dot{\beta}}\Phi_{(\alpha(s-3)\beta)(\dot{\alpha}(s-3)\dot{\beta})}
			\\&
			-
			\frac{2s^2}{2s-1}  
			\partial^{\beta\dot{\beta}}\partial^{\gamma\dot{\gamma}} \Phi_{(\alpha(s-2)\beta\gamma)(\dot{\alpha}(s-2)\dot{\beta}\dot{\gamma})}.
		\end{split}
	\end{equation}
\end{subequations}
Using these definitions, one can show that  on the spin-1 mass shell
\begin{equation}
	\mathcal{L}^{CC}_{\text{fake}}
	\approx
	\Phi^{\alpha(3)\dot{\alpha}(3)}	\mathbf{J}_{\alpha(3)\dot{\alpha}(3)}^{C C}
	+
	\Phi^{\alpha\dot{\alpha}}	\mathbf{T}_{\alpha\dot{\alpha}}^{C C}.
\end{equation}	
We conclude that no non-trivial vertex of this type exists; the constructed vertex can be eliminated by a local field redefinition.

\medskip

\smallskip

\noindent \textbf{B.  $J \sim C \bar{C}$}

\smallskip

Let us now consider a vertex  with 5 derivatives, constructed from spin-1 Weyl tensors of different chirality. The general ansatz for the traceless spin-3 current is given by:
\begin{equation}
	\mathbf{J}_{\alpha(3)\dot{\alpha}(3)}^{C \bar{C}} 
	=
	 a_1  (\partial C_1)_{\alpha(2)\beta \dot{\beta}} 
	 (\partial^2 \bar{C}_1)_{\alpha\dot{\alpha}(3)}^{\beta\dot{\beta}}
	 +
	 a_2
	 (\partial^2 C_1)_{\alpha(3)\beta\dot{\alpha}\dot{\beta}} 
	 (\partial \bar{C}_1)_{\dot{\alpha}(2)}^{\beta\dot{\beta}}.
\end{equation}	
From the spin-1 curvatures and three derivatives, it is impossible to construct the  trace for the spin-3 current:
\begin{equation}
	\mathbf{T}^{C \bar{C}}_{\alpha\dot{\alpha}}
	=
	0.
\end{equation}	

Then the conservation condition takes the form
\begin{equation}
	\partial^{\rho\dot{\rho}} \mathbf{J}^{C \bar{C}}_{\rho\alpha(2)\dot{\rho} \dot{\alpha}(2)}
	\approx
	\frac{1}{3} \left[ a_1 +a_2   \right] (\partial^2 C_1)_{\alpha(2)\beta(2) \dot{\beta}(2)} 
(\partial^2 \bar{C}_1)_{\dot{\alpha}(2)}^{\beta(2)\dot{\beta}(2)}
\end{equation}	
and leads to the equality
\begin{equation}
	a_2 = -a_1.
\end{equation}	

\medskip
In this case, there also exists a fake vertex that corresponds to the constructed current
\begin{equation}\label{eq: fake 5-2}
\mathcal{L}^{C\bar{C}}_{\text{fake}}
\sim	\mathcal{R}^{\alpha(3)\dot{\alpha}(3)}_{(s=3)}
	(C_1)_{\alpha(2)} (\partial \bar{C}_1)_{\alpha\dot{\alpha}(3)}
	\sim
	\Phi^{\alpha(3)\dot{\alpha}(3)}	\mathbf{J}_{\alpha(3)\dot{\alpha}(3)}^{C \bar{C}}. 
\end{equation}	
As a result, we conclude that it is impossible to construct a non-trivial vertex with five derivatives for the interaction of a spin-$3$ field with a single spin-$1$ field. This differs from the situation for the vertex with the minimal number of derivatives (three derivatives), where one can construct a parity-odd vertex, see, e.g., \cite{Zaigraev:2026jvu}.

\newpage

\noindent \textbf{C.  $J \sim C B$}

\smallskip

Let us consider two different spin-$1$ fields with Weyl curvatures, denoted as $C_1$ and $B_1$. The general ansatz with $5$ derivatives for the traceless current is given by
\begin{equation}
	\begin{split}
		\mathbf{J}^{CB}_{\alpha(3)\dot{\alpha}(3)}
		=&
		\;
		a_1 \left(C_1 \right)_{\beta(2)} \left( \partial^3 B_1 \right)^{\beta(2)}_{\alpha(3)\dot{\alpha}(3)}
		+
		a_2
		\left( \partial C_1 \right)_{\alpha\beta(2)\dot{\alpha}} (\partial^2B_1 )^{\beta(2)}_{\alpha(2)\dot{\alpha}(2)}
		\\
		&
		+
		a_3 	\left( \partial^2 C_1 \right)_{\alpha(2)\beta(2)\dot{\alpha}(2)} (\partial B_1 )^{\beta(2)}_{\alpha\dot{\alpha}}
		+
		a_4	\left( \partial^3 C_1 \right)_{\alpha(3)\beta(2)\dot{\alpha}(3)} ( B_1 )^{\beta(2)},
	\end{split}
\end{equation}	
where $a_i$ are arbitrary complex coefficients.
We see that, in contrast to \eqref{eq: CC with one spin 1}, in this case there are four independent complex coefficients.
 The corresponding ansatz for the current trace takes the form
\begin{equation}
	\mathbf{T}^{CB}_{\alpha\dot{\alpha}}
	=
	b_1 	(\partial C_1)_{\beta(3)\dot{\beta}} (\partial^2 B_1)^{\beta(3)\dot{\beta}}_{\alpha\dot{\alpha}}
	+
	b_2 	(\partial^2 C_1)_{\alpha\dot{\alpha}\beta(3)\dot{\beta}} (\partial B_1)^{\beta(3)\dot{\beta}}.
\end{equation}	
Substituting these expressions into the higher-spin current conservation equation and using the free on-shell equations of motion for both spin-$1$ fields ($\partial^{\alpha\dot{\alpha}}C_{\alpha(2)} \approx 0$ and $\partial^{\alpha\dot{\alpha}}B_{\alpha(2)} \approx 0$), we obtain the following system of linear algebraic equations:
\begin{equation}{\label{eq: system CB}}
	\begin{cases}
	a_1 + \frac{1}{9} a_2 + b_1 = 0,
	\\
	\frac{4}{9} a_2 + \frac{4}{9} a_3 + b_1 + b_2 = 0,  \\
	\frac{1}{9} a_3 + a_4 + b_2 = 0. 
	\end{cases}
\end{equation}
Since the system contains three equations for six coefficients, the general solution possesses $3$ free parameters. Choosing $a_1$, $b_1$, and $b_2$ as the independent constants, the solution reads:
\begin{equation}\label{eq: solution CB}
	\begin{aligned}
		a_2 &= -9 a_1 - 9 b_1, \\
		a_3 &= 9 a_1 + \frac{27}{4} b_1 - \frac{9}{4} b_2, \\
		a_4 &= -a_1 - \frac{3}{4} b_1 - \frac{3}{4} b_2.
	\end{aligned}
\end{equation}
We see that in total, there are three complex currents, but there are only two complex fake vertices\footnote{Note that the expressions multiplying the tensor $\mathcal{R}^{\alpha\dot{\alpha}}_{(s=3)}$ correspond to conserved spin-1 currents with three derivatives, which corresponds to the 3-derivative $(1,1,1)$ vertex.}:
\begin{subequations}
\begin{equation}\label{eq: fake 5-2 pr}
	\mathcal{L}^{CB}_{\text{fake}-1}
	\sim
	\mathcal{R}^{\alpha\dot{\alpha}}_{(s=3)} (\partial C_1)_{\alpha\beta(2)\dot{\alpha}} (B_1)^{\beta(2)},
\end{equation}
\begin{equation}\label{eq: fake 5-2 pr2}
	\mathcal{L}^{CB}_{\text{fake}-2}
	\sim
	\mathcal{R}^{\alpha\dot{\alpha}}_{(s=3)} (\partial B_1)_{\alpha\beta(2)\dot{\alpha}} (C_1)^{\beta(2)}.
\end{equation}
\end{subequations}
This means that only one complex current remains, corresponding to non-trivial $(3,1,1)$ interactions.

\smallskip

For example, by setting the trace parameters to zero ($b_1 = b_2 = 0$) and choosing $a_1 = 1$, one uncovers a strictly conserved traceless current
\begin{equation}
	\begin{split}
		\mathbf{J}^{CB}_{\alpha(3)\dot{\alpha}(3)}
		=&
		 \left(C_1 \right)_{\beta(2)} \left( \partial^3 B_1 \right)^{\beta(2)}_{\alpha(3)\dot{\alpha}(3)}
		-
		9
		\left( \partial C_1 \right)_{\alpha\beta(2)\dot{\alpha}} (\partial^2B_1 )^{\beta(2)}_{\alpha(2)\dot{\alpha}(2)}
		\\
		&
		+
	9	\left( \partial^2 C_1 \right)_{\alpha(2)\beta(2)\dot{\alpha}(2)} (\partial B_1 )^{\beta(2)}_{\alpha\dot{\alpha}}
		-
			\left( \partial^3 C_1 \right)_{\alpha(3)\beta(2)\dot{\alpha}(3)} ( B_1 )^{\beta(2)}.
	\end{split}
\end{equation}	
The real and imaginary parts of this spin-3 current  correspond to physically non-trivial parity-even and parity-odd cubic Born-Infeld interactions that do not vanish on the spin-$3$ mass shell.

\smallskip

The consideration of the current constructed from Weyl tensors of different chirality $J^{C\bar{B}}$ is completely analogous to that of the current $J^{C\bar{C}}$  and leads to the conclusion that such vertices do not exist.

\smallskip

To summarize, the analysis of the $(3,1,1)$ vertex shows that:
(i) a non-trivial Born--Infeld vertex requires two distinct fields of
the same chirality; 
(ii) for a single field or for fields of opposite
chirality only fake vertices exist; 
(iii) the resulting non-trivial
vertex has a parity-even and a parity-odd part.
 The general
construction of Section~\ref{sec: general current} will reproduce these
statements for arbitrary spins.

\section{General Born-Infeld current}\label{sec: general current}

Now we turn to the generalization of the illustrated construction to the case of arbitrary spins.
The general gauge-invariant ansatz for the higher-spin current and the current trace corresponding to the $(s, s_1,s_2)$ vertex  with $s+s_1+s_2$ derivatives are given by:
\begin{subequations}
\begin{equation}\label{eq: mathbf J}
	\mathbf{J}_{\alpha(s)\dot{\alpha}(s)}
	=
	\sum_{p=0}^{s- (s_1-s_2)}
	a_p \left(\partial^p C_{s_1}\right)_{\alpha(s_1-s_2+p) \beta(s_1+s_2) \dot{\alpha}(p)}
	\left( \partial^{s-p} C_{s_2} \right)^{\beta(s_1+s_2)}_{\alpha(s+s_2-s_1-p)\dot{\alpha}(s-p)},
\end{equation}	
\begin{equation}
	\begin{split}
	\mathbf{T}_{\alpha(s-2)\dot{\alpha}(s-2)}
	=
	\sum_{p=0}^{s - (s_1-s_2)-2}
	b_p
	& \left(\partial^{p+1} C_{s_1}\right)_{\alpha(s_1-s_2+p) \beta(s_1+s_2+1) \dot{\alpha}(p) \dot{\beta}}
	\\
&\times	\left( \partial^{s-p-1} C_{s_2} \right)^{\beta(s_1+s_2+1)\dot{\beta}}_{\alpha(s+s_2-s_1-p-2)\dot{\alpha}(s-p-2)}.
	\end{split}
\end{equation}	
\end{subequations}
Here $a_p$ and $b_p$ are arbitrary complex coefficients.  We assume that $s\geq s_1 \geq s_2$ and the upper limits in the sums are determined by the requirement that the number of $\alpha$-indices in the second multiplier must be non-negative.

Substituting the ansatz into the higher-spin current conservation equation \eqref{eq: cons equation}, we obtain the recurrence relations\footnote{It is interesting to note that  the coefficients in the recurrence relations after the change $s_2\to -s_2$ coincide with those in the recurrence relation for  Abelian higher-spin currents  \cite{Zaigraev:2026xzr}. In the case of the supersymmetric currents considered below, there is no such notable coincidence.}
\begin{equation}\label{eq: BI recurrence}
	a_p\cdot  \frac{s-s_1+s_2-p}{s}
	\cdot  \frac{s-p}{s}
	+
	a_{p+1}\cdot \frac{s_1-s_2+p+1}{s} \cdot \frac{p+1}{s}
	+
	b_p+b_{p-1}=0,
\end{equation}
with the boundary conditions
\begin{equation}
	\begin{split}
	&a_p=0 \qquad \text{for}  \qquad p<0 \quad \text{or} \quad p>s - (s_1-s_2),
	\\
	&b_p=0\qquad \text{for}  \qquad p<0 \quad \text{or}  \quad  p > s-  (s_1-s_2) -2.
	\end{split}
\end{equation}
The solution of the resulting equations depends on the choice of complex coefficients $b_p$. 

\medskip

$\bullet$ For the traceless case $b_p=0$, the solution is
\begin{equation}\label{eq: BI traceless}
		a_p^{(0)}
		=
		(-1)^p\,
		\frac{\binom{s}{s_1-s_2+p} \binom{s}{p}}{\binom{s}{s_1-s_2}},
		\qquad
		p=0,1,\dots,s - (s_1-s_2).
\end{equation}

\smallskip

If we consider the case $s_1=s_2$ and only one spin-$s_1$ field, structures with $p$ and $s-p$ coincide:
\begin{equation}
	\begin{split}
		K_p = \left(\partial^p C_{s_1}\right)_{\alpha(p) \beta(2s_1) \dot{\alpha}(p)}
		\left( \partial^{s-p} C_{s_1} \right)^{\beta(2s_1)}_{\alpha(s-p)\dot{\alpha}(s-p)},
		\\
		K_{s-p} 
		=
		\left(\partial^{s-p} C_{s_1}\right)_{\alpha(s-p) \beta(2s_1) \dot{\alpha}(s-p)}
		\left( \partial^{p} C_{s_1} \right)^{\beta(2s_1)}_{\alpha(p)\dot{\alpha}(p)},
	\end{split}
\end{equation}	
so that the traceless current  reduces to
\begin{equation}\label{eq: odd gen}
	\mathbf{J}^{(0)}_{\alpha(s)\dot{\alpha}(s)}
	=
	\sum_{p=0}^{s} a_p^{(0)} K_p
	=
	\sum_{p=0}^{\lfloor s/2\rfloor}
	\bigl(a_p^{(0)} + a_{s-p}^{(0)}\bigr) K_p
	=
	\sum_{p=0}^{\lfloor s/2\rfloor}
	(-1)^p \binom{s}{p}^2
	\bigl(1+(-1)^s\bigr) K_p.
\end{equation}
Hence $\mathbf{J}^{(0)}$ vanishes identically for odd $s$, while for even $s$ it is non-zero.
This is a generalization of the particular statement that we encountered in the previous section \ref{sec: 2}.
In the odd $s$ case there are no traceless conserved currents; the conserved currents are necessarily traceful (which, however, corresponds to a fake vertices).

\medskip

$\bullet$ In the general case with non-vanishing trace, the solution of \eqref{eq: BI recurrence} is
\begin{equation}\label{eq: BI general}
		a_p
		=
		a_p^{(0)}
		\left[
		\frac{a_0}{a_0^{(0)}}
		+
		s^2
		\sum_{j=0}^{p-1}
		\frac{b_j+b_{j-1}}
		{(s-s_1+s_2-j)(s-j)\,a_j^{(0)}}
		\right],
			\qquad
		p=0,1,\dots,s - (s_1-s_2),
\end{equation}
where $a_p^{(0)}$ is given by \eqref{eq: BI traceless}. The coefficients $b_p$ remain arbitrary and parameterize the trace part of the current.

In total there are $s - (s_1-s_2)$ independent complex conserved currents. In order to single out from them the currents that correspond to non-trivial interactions, let us consider fake vertices.

\subsection*{Fake vertices}

To construct fake vertices, we will use gauge-invariant Fronsdal curvatures for spin-$s$ given in \eqref{eq: R curv}. $\mathcal{R}$-curvatures contain two derivatives, and the general form of such vertices is as follows\footnote{It is impossible to construct a vertex using the tensor $\mathcal{R}^{\alpha(s)\dot{\alpha}(s)}$, since the dotted $s-2$ indices arising from the $s-2$ derivatives cannot be contracted with the $s$ indices of the curvature. }:
 \begin{equation}
\mathcal{L}^{(s, s_1, s_2)}_{\text{fake}} \sim 
\mathcal{R}_s^{\alpha(s-2)\dot{\alpha}(s-2)}
	\cdot \mathbf{J}_{\alpha(s-2)\dot{\alpha}(s-2)}.
\end{equation}
Here $\mathbf{J}_{\alpha(s-2)\dot{\alpha}(s-2)}$  is given by eq. \eqref{eq: mathbf J} and parametrized by $s-(s_1-s_2)-1$ arbitrary complex coefficients.  By construction this current contains $s-2$ derivatives.

\smallskip 

As a result, we see that only one non-trivial conserved complex current remains, which gives rise to a non-trivial interactions. Its real part corresponds to a parity-even interaction, while its imaginary part corresponds to a parity-odd interaction. 
It should be emphasized that this statement is understood up to the addition of fake currents: different representatives of the same equivalence class, related by local field redefinitions, describe the same physical vertex. The choice of the traceless representative is a matter of convenience and does not affect the physical content.

\subsection*{Case $s \leq s_1+s_2$ and manifestly gauge-invariant representation}

In the case $s \leq s_1+s_2$, the vertex can be constructed exclusively in terms of the higher-spin Weyl tensors, as shown in \eqref{eq: C3}. Let us show that such a vertex corresponds to the general traceless current, given by \eqref{eq: BI traceless}.

To this end, let us derive a useful identity which is satisfied on the spin-$s_1$ and spin-$s_2$ mass shell.
 Let us denote $s_1+s_2-s = k$, then
 \begin{subequations}\label{eq: ident W and currents}
\begin{equation}
	\begin{split}
	I  &= 	(\partial^s)_{\dot{\alpha}(s)}^{\alpha(s)}
			\left\{
	(C_{s_1})_{\alpha(2s_1-k)\rho(k)} (C_{s_2})^{\rho(k)}_{\alpha(2s_2-k)}
		\right\}
	\\&	=
		\sum_{p=0}^s \binom{s}{p} 
		\left\{
		\left(\partial^p\right)_{\dot{\alpha}(p)}^{\alpha(p)}  	(C_{s_1})_{\alpha(2s_1-k)\rho(k)}  \right\}
		\left\{
		\left(\partial^{s-p}\right)_{\dot{\alpha}(s-p)}^{\alpha(s-p)}  	 (C_{s_2})^{\rho(k)}_{\alpha(2s_2-k)}  \right\}
	\end{split}
\end{equation}	
Then we use the conservation equations \eqref{eq: weyl cons} for the higher-spin Weyl tensors, which hold on the mass shell of spin-$s_1$ and spin-$s_2$ fields:
\begin{equation}
	\begin{split}
		I &\approx
		\sum_{p=0}^s \binom{s}{p} 
	\frac{2s_2-k}{2(s_1+s_2-k)}  \dots  \frac{2s_2-k-p+1}{2(s_1+s_2-k)-p+1} 
	\\
	&	\qquad\qquad\quad
		\frac{2s_1-k}{2(s_1+s_2-k)-p}
		\dots
		\frac{2s_1-k-s+p+1}{2(s_1+s_2-k)- s +1}
		\\&
		\qquad \qquad\quad\left\{
		\left(\partial^p\right)_{\dot{\alpha}(p)}^{\gamma(p)}  
		\left( C_{s_1} \right)_{\alpha(2s_1-k-s+p)\beta(s-p)\rho(k)}   \right\}
		\left\{
		\left(\partial^{s-p}\right)_{\dot{\alpha}(s-p)}^{\beta(s-p)}  	\left( C_{s_2} \right)^{\rho(k)}_{\alpha(2s_2-k-p)\gamma(p)}   \right\}.
	\end{split}
\end{equation}	
Extracting the binomial coefficients, we arrive at the equality
\begin{equation}
	\begin{split}
	I 	&\approx
		\sum_{p=0}^s 
		(-1)^p
		\frac{		\binom{s}{p}  \binom{s}{s_1-s_2+p}}{ \binom{2s}{s_1-s_2+s}}
		(\partial^p C_{s_1})_{\alpha(s_1-s_2+p)\,\beta(s_1+s_2)\,\dot{\alpha}(p)}
		(\partial^{s-p} C_{s_2})^{\beta(s_1+s_2)}_{\alpha(s-s_1+ s_2 -p)\,\dot{\alpha}(s-p)}
		\\
		&=
		\frac{\binom{s}{s_1-s_2}}{\binom{2s}{s_1-s_2+s}} \mathbf{J}^{0}_{\alpha(s)\dot{\alpha}(s)}.
	\end{split}
\end{equation}	
 \end{subequations}
The sign  $(-1)^p$
in this equality arises from lowering the $p$ contracted undotted indices.

\medskip

Thus, we see that, up to a numerical factor, in the case $s\leq s_1 +s_2$ the following equality between the vertices holds:
\begin{equation}
	\mathcal{L}_{(s,s_1,s_2)} \sim	C_s^{\alpha(2s)} (C_{s_1})_{\alpha(2s_1-k)\beta(k)} (C_{s_2})^{\beta(k)}_{\alpha(2s_2-k)}
	\sim 
	\Phi^{\alpha(s)\dot{\alpha}(s)} \mathbf{J}^{0}_{\alpha(s)\dot{\alpha}(s)}.
\end{equation}	
The form of the vertex in terms of the higher-spin Weyl tensors is manifestly gauge-invariant.

 \section{$\mathcal{N}=1$ Born-Infeld interactions}\label{sec: N=1 BI}
 
 Let us now consider the generalization to the case of $\mathcal{N}=1$ supersymmetry.
 We will consider a cubic vertex
 $(s, Y_1, Y_2)$ between arbitrary massless  superspins $Y_1, Y_2$  and a massless integer superspin $s$. 
 In the special case of identical superspins $Y_1=Y_2=Y$ such interactions  were considered in the paper  \cite{Gates:2019cnl}.
 
 \smallskip
   
 Let us recall  that cubic interactions of this type can be constructed using conserved $\mathcal{N}=1$ supercurrents and are given by
 (see, e.g., \cite{Buchbinder:2018wzq, Gates:2019cnl})
 \begin{equation}
 	S_{(s, Y_1, Y_2)}= \int d^8z \left\{ 
 	\left( \Psi^{\alpha(s)\dot{\alpha}(s-1)} \mathcal{J}_{\alpha(s)\dot{\alpha}(s-1)}
 	+
 	c.c. \right)
 	+
 	V^{\alpha(s-1)\dot{\alpha}(s-1)} \mathcal{T}_{{\alpha(s-1)\dot{\alpha}(s-1)}}
 	\right\},
 \end{equation}	
 where $\Psi$ and $V$ are higher-spin prepotentials of integer superspin.
 For the interactions to be consistent, the supercurrent and supertrace must be gauge invariant and satisfy conservation equations on the mass shell:
 \begin{equation}\label{eq: cons full}
 	\begin{cases}
 		\mathcal{D}^2 \mathcal{J}_{\alpha(s)\dot{\alpha}(s-1)} \approx \frac{1}{s!} \mathcal{D}_{\alpha} \mathcal{T}_{{\alpha(s-1)\dot{\alpha}(s-1)}},
 		\\
 		\bar{\mathcal{D}}^{\dot{\alpha}} \mathcal{J}_{\alpha(s)\dot{\alpha}(s-1)} \approx 0.
 	\end{cases}	
 \end{equation}	
 We will strengthen these equations and require $\mathcal{T} = 0$ and the fulfillment of
 \begin{equation}\label{eq: conf cons}
 	 	\begin{cases}
 		\mathcal{D}^\alpha \mathcal{J}_{\alpha(s)\dot{\alpha}(s-1)} \approx 0,
 		\\
 		\bar{\mathcal{D}}^{\dot{\alpha}} \mathcal{J}_{\alpha(s)\dot{\alpha}(s-1)} \approx 0.
 	\end{cases}	
 \end{equation}	
 The strengthening of the conservation equations to the form \eqref{eq: conf cons} is motivated as follows. Solutions of the full system \eqref{eq: cons full} with a non-trivial supertrace correspond to fake interactions, i.e. vertices that can be removed by local field redefinitions. Imposing \eqref{eq: conf cons} selects a distinguished representative among the equivalence classes of conserved supercurrents, while leaving the physical (non-fake) vertices intact. In this sense, \eqref{eq: conf cons} may be viewed as the conservation equations for coupling to a conformal integer-superspin-s multiplet  (see discussion in \cite{Buchbinder:2018wzq, Gates:2019cnl}).

 \medskip
 
 The main objects for constructing conserved supercurrents are gauge-invariant higher-spin Weyl $\mathcal{N}=1$ supertensors, which for integer superspins are given by (see, e.g. \cite{BK})
 \begin{subequations}
 \begin{equation}\label{eq: int superspin W}
 	\left( \mathcal{W}_{s} \right)_{\alpha(2s)} = \bar{\mathcal{D}}^2 \mathcal{D}_\alpha (\partial^{s-1})_{\alpha(s-1)}^{\dot{\alpha}(s-1)} \Psi_{\alpha(s)\dot{\alpha}(s-1)}.
 \end{equation}	
 Similarly, for half-integer superspin, the corresponding supertensors are given by
 \begin{equation}
 	\left( \mathcal{W}_{s+\frac{1}{2}} \right)_{\alpha(2s+1)} = \bar{\mathcal{D}}^2 \mathcal{D}_\alpha (\partial^{s})_{\alpha(s)}^{\dot{\alpha}(s)} H_{\alpha(s)\dot{\alpha}(s)}.
 \end{equation}	
 \end{subequations}
 In all formulas, we assume full symmetrization over undotted indices.
 
 \smallskip
 
 These supertensors satisfy the chirality condition ($ 	\bar{\mathcal{D}}_{\dot{\beta}} \mathcal{W}_{\alpha(2Y)} = 0$) and useful conservation laws on the mass shell, which we will employ:
 \begin{equation}\label{eq: N=1 eom}
 	\mathcal{D}^\beta \mathcal{W}_{\beta\alpha(2Y-1)} \approx 0,
 	\qquad
 	\partial_{\dot{\beta}}^\beta \mathcal{W}_{\beta\alpha(2Y-1)} \approx 0,
 	\qquad
 	\mathcal{D}^2 \mathcal{W}_{\alpha(2Y)} \approx 0.
 \end{equation}	
 
 \smallskip
 
We consider the following general ansatz for the $\mathcal{N}=1$ higher-spin gauge-invariant supercurrent constructed from Weyl tensors of the same chirality\footnote{Such an ansatz is possible only if the superspins $Y_1$ and $Y_2$ are either both integer or both half-integer (i.e. $|Y_1-Y_2| \in \mathbb{Z}$), in full agreement with the Metsaev classification.}:
  \begin{equation}
 	\begin{split}
 	\mathcal{J}_{\alpha(s)\dot{\alpha}(s-1)}
 	=
 	&
 	\sum^{s-(Y_1-Y_2)-1}_{p=0} a_p
 	 \left(\mathcal{D} \partial^p \mathcal{W}_{Y_1}\right)_{\alpha(Y_1-Y_2+p+1)\beta(Y_1+Y_2)\dot{\alpha}(p)} 
 	 \left(
 	 \partial^{s-p-1} \mathcal{W}_{Y_2}
 	 \right)^{\beta(Y_1+Y_2)}_{\alpha(s-(Y_1-Y_2)-p-1) \dot{\alpha}(s-p-1)}
 	 \\
 	 &
 	  +
 	  	\sum_{p=0}^{s-(Y_1-Y_2)} b_p
 	 \left(\partial^p \mathcal{W}_{Y_1}\right)_{\alpha(Y_1-Y_2+p)\beta(Y_1+Y_2)\dot{\alpha}(p)} 
 	 \left(
 	 \mathcal{D} 
 	 \partial^{s-p-1} \mathcal{W}_{Y_2}
 	 \right)^{\beta(Y_1+Y_2)}_{\alpha(s-(Y_1-Y_2)-p) \dot{\alpha}(s-p-1)}.
 	 \end{split}
 \end{equation}	
 Here $a_p$ and $b_p$ are arbitrary complex coefficients.
 In all factors, we assume full symmetrization over all indices; otherwise, the corresponding factor would vanish on the mass shell, according to~\eqref{eq: N=1 eom}.
 The upper limits in the sums follow from the requirement that the number of 
 $\alpha$-indices in the second factors be non-negative.
 
For compactness of further expressions, we introduce $\Delta = Y_1-Y_2$. The supercurrent conservation equations \eqref{eq: conf cons} lead to a system of recurrence relations for the coefficients:
 \begin{subequations}
 \begin{equation}
 	\mathcal{D}^{\alpha}
 		\mathcal{J}_{\alpha(s)\dot{\alpha}(s-1)}
 		\approx 0
 		\qquad
 		\Rightarrow
 		\qquad
 		(-1)^{2Y_1 } a_p \cdot \frac{\Delta +p+1}{s}
 		+
 		b_p \cdot \frac{s-\Delta -p}{s} = 0,
 \end{equation}	
 \begin{equation}\label{eq: N=1 eq a and b}
 	\bar{\mathcal{D}}^{\dot{\alpha}}
 		\mathcal{J}_{\alpha(s)\dot{\alpha}(s-1)}
 	\approx 0
 	\qquad
 	\Rightarrow
 	\qquad
 -  a_p \cdot \frac{s -p-1}{s-1}
  +
(-1)^{2Y_1}   b_{p+1} \cdot \frac{p+1}{s-1} = 0.
 \end{equation}	
 \end{subequations}
  The general solution of these equations is given by
\begin{subequations}
	\begin{equation}
		a_p = c \cdot (-1)^p \cdot \binom{s-1}{p} \binom{s}{\Delta+p+1},
		\qquad 0 \le p \le s-\Delta-1,
	\end{equation}
	\begin{equation}
		b_p = c\cdot (-1)^{2Y_1 +1 } \cdot (-1)^p \cdot \binom{s-1}{p} \binom{s}{\Delta+p},
		\qquad 0 \le p \le s-\Delta,
	\end{equation}
\end{subequations}
  where $c$ is an overall complex normalization constant.
  The real and imaginary parts of the constructed supercurrent correspond to parity-even and parity-odd consistent vertices, respectively.

  \medskip
  
    Let us discuss the obtained result.
    
    \smallskip
 
 \noindent $\bullet$ \textbf{Equal superspins} 
 
     \smallskip
 
 When the higher-spin supercurrent is constructed from Weyl supertensor of the same superspin ($Y_1 = Y_2$),  the recurrence relations simplify significantly, and the coefficients $a_p$ and $b_p$ can be written as straightforward products of binomial coefficients:
 \begin{subequations}\label{eq:equal_spins}
 	\begin{equation}
 		a_p = c \cdot (-1)^p \cdot \dbinom{s-1}{p} \dbinom{s}{p+1}, 
 		\qquad
 		 0 \le p \le s-1
 	\end{equation}
 	\begin{equation}
 		b_p = c \cdot  (-1)^{2Y_1 +1} \cdot (-1)^p \cdot \dbinom{s-1}{p} \dbinom{s}{p}, \qquad 0 \le p \le s-1.
 	\end{equation}
 \end{subequations}
Upon identifying the superspin-$Y_1$
and superspin-$Y_2$ superfields, the expression for the supercurrent becomes (we use relation $a_p =  (-1)^{2Y_1 } \cdot (-1)^{s} b_{s-p-1} $)
  \begin{equation}
 	\begin{split}
 		\mathcal{J}_{\alpha(s)\dot{\alpha}(s-1)}
 		=
 		&
 		\sum^{s-1}_{p=0}a_p  \left\{ 
 		1  +
 		(-1)^{s} 
 		 \right\}
 		\left(\mathcal{D} \partial^p \mathcal{W}_{Y_1}\right)_{\alpha(p+1)\beta(2Y_1)} 
 		\left(
 		\partial^{s-p-1} \mathcal{W}_{Y_1}
 		\right)^{\beta(2Y_1)}_{\alpha(s-p-1)}.
 	\end{split}
 \end{equation}	
 We see that for odd values of 
 $s$, the resulting supercurrent vanishes identically.
 This statement is a supersymmetric version of the results analyzed in the previous section \ref{eq: general current}, see eq. \eqref{eq: odd gen}.
 The obtained supercurrent is in full agreement with earlier results of Gates and Koutrolikos
 \cite{Gates:2019cnl}.

    \smallskip
 
 \noindent $\bullet$ \textbf{ $s \leq Y_1 + Y_2$  case} 
 
 \smallskip
 
 In the case $s \leq Y_1 + Y_2$ one can represent the interaction in a manifestly gauge-invariant form, as the integral over $\mathcal{N}=1$ superspace
 \begin{equation}
 		S^{\text{chiral}}_{(s, Y_1, Y_2)} 
 		=
 		\int d^4x d^2 \theta
 		\;
 		 \mathcal{W}_s^{\alpha(2s)}
 		\left( \mathcal{W}_{Y_1} \right)_{\alpha(Y_1-Y_2+s)\beta(Y_1+Y_2-s)} 
 		\left( \mathcal{W}_{Y_2} \right)^{\beta(Y_1+Y_2-s)}_{\alpha(s-Y_1+Y_2 )}. 
 \end{equation}	
 Actually, using the definition \eqref{eq: int superspin W} of $\mathcal{W}_s$, the relations $d^8z = d^4x d^2\theta \bar{\mathcal{D}}^2 $ and the generalization of identity \eqref{eq: ident W and currents}, we obtain up to a numerical coefficient
 \begin{equation}
 S^{\;\text{chiral}}_{(s, Y_1, Y_2)} 
 \sim
 \int d^8 z \Psi^{\alpha(s)\dot{\alpha}(s-1)} \mathcal{J}_{\alpha(s)\dot{\alpha}(s-1)}.
 \end{equation}	
 Thus, the supercurrent found corresponds to all possible interactions of $(s, Y_1, Y_2)$ type with integer superspin $s$, including those constructed from gauge-invariant Weyl supertensors.
 
  \smallskip
 
 \noindent $\bullet$ \textbf{ Supercurrents with non-trivial supertrace} 
 
 \smallskip
 
 A more general set of supercurrents with a non-trivial supertrace can be constructed, as was shown for the case of identical superspins in the paper \cite{Gates:2019cnl}. However, all these supercurrents correspond to fake vertices, analogous to the component analysis in the previous section. 
 
     \smallskip
 
 \noindent $\bullet$ \textbf{ Half-integer superspin $s$ case} 
 
 \smallskip
 
 For half-integer superspins, it is impossible to construct a consistent cubic vertex and corresponding supercurrent from higher-spin Weyl supertensors of the same chirality. This can be directly verified by considering the most general ansatz for the real supercurrent and imposing the conservation conditions. Alternatively, this can be seen by noting that it is impossible to construct a Lorentz-invariant scalar from three Weyl supertensors with an odd number of indices. These observations are consistent with the Metsaev classification of $\mathcal{N}=1, 4D$ cubic vertices \cite{Metsaev:2019dqt}.
 
 \section{$\mathcal{N}=2$ Born-Infeld interactions}\label{sec: N=2 gen}
 
 Now we turn to the generalization to the case of extended $\mathcal{N}=2$ supersymmetry.
 We use standard notation, following \cite{18}.
Using Mezincescu-type higher-spin prepotentials $\Psi^{-\alpha(s-1)\dot{\alpha}(s-2)}$ and $\bar{\Psi}^{-\alpha(s-2)\dot{\alpha}(s-1)}$\footnote{Mezincescu-type prepotentials definition and their relation to unconstrained higher-spin analytic prepotentials are described in detail in the works \cite{Buchbinder:2022vra, Zaigraev:2026jvu}. For $\mathcal{N}=2$ supergravity ($s=2$) these prepotentials were introduced by Zupnik \cite{Zupnik:1998td} and correspond to harmonic superspace generalization of Gates-Siegel prepotentials \cite{Gates:1981qq}. }, one can construct cubic $(\mathbf{s}, \mathbf{s_1}, \mathbf{s_2})$\footnote{We denote by $\mathbf{s}$ a higher-spin $\mathcal{N}=2$ supermultiplet $\left\{ s, 2\times \left(s-\frac{1}{2}\right), s-1\right \}$. In terms of $\mathcal{N}=1$ supermultiplets, for integer highest spin we can represent such a supermultiplet as $\{\Phi_s, \Psi_{s-1}\}$, and for half-integer highest spin as $\{\Phi_s, \Psi_s\}$.} interactions of the following form \cite{Zaigraev:2024ryg}:
 \begin{equation}\label{eq: Psi inter}
 	S_{(\mathbf{s}, \mathbf{s_1}, \mathbf{s_2})} = \int d^4x d^8\theta du \; \left(\Psi^{-\alpha(s-1)\dot{\alpha}(s-2)} \mathcal{J}^+_{\alpha(s-1)\dot{\alpha}(s-2)}
 	-
 	\bar{\Psi}^{-\alpha(s-2)\dot{\alpha}(s-1)} \bar{\mathcal{J}}^+_{\alpha(s-2)\dot{\alpha}(s-1)}
 	\right),
 \end{equation}	
 where, for consistency, $\mathcal{N}=2$ supercurrents must satisfy the on-shell conservation laws:
 \begin{equation}\label{eq: conservation supercurrents}
 	\begin{cases}
 		\mathcal{D}^{++} \mathcal{J}^+_{\alpha(s-1)\dot{\alpha}(s-2)} \approx 0,
 		\\
 		\mathcal{D}^{+\alpha} \mathcal{J}^+_{\alpha(s-1)\dot{\alpha}(s-2)} + \bar{\mathcal{D}}^{+\dot{\alpha}} \bar{\mathcal{J}}^+_{\alpha(s-2)\dot{\alpha}(s-1)} \approx 0,
 		\\
 		\mathcal{D}^+_{(\beta} \mathcal{J}^+_{\alpha(s-1))\dot{\alpha}(s-2)} \approx 0,
 		\\
 		\bar{\mathcal{D}}^+_{(\dot{\alpha}} \mathcal{J}^+_{\alpha(s-1)\dot{\alpha}(s-2))} - \mathcal{D}^+_{(\alpha} \bar{\mathcal{J}}^+_{\alpha(s-2))\dot{\alpha}(s-1)} \approx 0,
 		\\
 		\bar{\mathcal{D}}^{+\dot{\alpha}} \mathcal{J}^+_{\alpha(s-1)\dot{\alpha}(s-2)} \approx 0. 
 	\end{cases}
 \end{equation}	
These equations admit a special class of solutions, which were used for the constructions
 of abelian interactions \cite{Zaigraev:2024ryg, Zaigraev:2026jvu, Zaigraev:2026xzr, Zaigraev:2025xqo}: 
 \begin{equation}\label{eq: J+ and J}
 	\begin{cases}
 		\mathcal{J}^+_{(\alpha_1\dots\alpha_{s-1})\dot{\alpha}(s-2)} = \mathcal{D}^+_{(\alpha_1} \mathcal{J}_{\alpha_2\dots\alpha_{s-1})\dot{\alpha}(s-2)},
 		\\
 		\bar{\mathcal{J}}^+_{\alpha(s-2)(\dot{\alpha}_1\dots\dot{\alpha}_{s-1})} = - \bar{\mathcal{D}}^+_{(\dot{\alpha}_1} \mathcal{J}_{\alpha(s-2)\dot{\alpha}_2\dots\dot{\alpha}_{s-1})}.
 	\end{cases}
 \end{equation}	
 
 Here we introduced the real \textit{principal $\mathcal{N}=2$ supercurrent} $\mathcal{J}_{\alpha(s-2)\dot{\alpha}(s-2)}$, which satisfies the system of equations:
 \begin{equation}\label{eq: primary supercurrent}
 	\begin{cases}
 		\widetilde{\mathcal{J}}_{\alpha(s-2)\dot{\alpha}(s-2)} = \mathcal{J}_{\alpha(s-2)\dot{\alpha}(s-2)},
 		\\
 		\mathcal{D}^{++} \mathcal{J}_{\alpha(s-2)\dot{\alpha}(s-2)} \approx 0,
 		\\
 		\mathcal{D}^{+\beta}\mathcal{J}_{(\beta\alpha(s-3))\dot{\alpha}(s-2)} \approx 0,
 		\\
 		\bar{\mathcal{D}}^{+\dot{\beta}} \mathcal{J}_{\alpha(s-2)(\dot{\beta}\dot{\alpha}(s-3))} \approx 0.
 	\end{cases}	
 \end{equation}	
 Here we assume $\mathbf{s}\geq 3$.
 In this form, these equations generalize the  $\mathcal{N}=1$ "conformal" conservation equations \eqref{eq: conf cons}.
 
 For the explicit construction of gauge-invariant 
$\mathcal{N}=2$ supercurrents, we will also use $\mathcal{N}=2$ Weyl supertensors $\mathcal{W}_{\alpha(2s-2)}$.
We will not present here their explicit construction in terms of analytic prepotentials (see details in \cite{Zaigraev:2026jvu, Ivanov:2024gjo, Ivanov:2025mld, Ivanov:2025gvs}); we will only give the main properties used below\footnote{At present, only the description of 
	$\mathcal{N}=2$ supermultiplets $\left\{ s, 2\times \left(s-\frac{1}{2}\right), s-1\right\}$ with the highest integer spin $s$ is known. However, on the mass shell, identical relations must hold for half-integer spins. Therefore, in what follows, we assume that the values of $s_1$ and $s_2$ can be either integer or half-integer.}.
First, off-shell the $\mathcal{N}=2$ higher-spin  Weyl supertensors satisfy the harmonic independence and chirality conditions
 \begin{equation}
	\mathcal{D}^{\pm\pm} \mathcal{W}_{\alpha(2s-2)} = 0,
\qquad
	\bar{\mathcal{D}}^\pm_{\dot{\alpha }}
	\mathcal{W}_{\alpha(2s-2)} = 0,
\end{equation}	
 On the mass shell,  higher-spin Weyl supertensors satisfy
 \begin{equation}
 	\mathcal{D}^{\pm\beta}	\mathcal{W}_{\beta\alpha(2s-3)} \approx 0
 \qquad
	\partial_{\dot{\alpha}}^\beta  \mathcal{W}_{\beta\alpha(2s-3)} \approx 0.
\end{equation} 
 
 Before proceeding to the construction of the $\mathcal{N}=2$ supercurrent, we note that if we construct a complex supercurrent satisfying the conservation equations, both its real and imaginary parts satisfy the reality condition; each of them gives a real interaction, corresponding to a parity-even and a parity-odd vertex, respectively.
So we consider the following ansatz for the complex 
$\mathcal{N}=2$ principal supercurrent,  constructed from the spin-$\mathbf{s_1}$ and spin-$\mathbf{s_2}$ $\mathcal{N}=2$ Weyl supertensors
 \begin{equation}
 	{\footnotesize
	\begin{split}
		&	\mathscr{J}_{\alpha(s-2)\dot{\alpha}(s-2)} 
		=
		\\
		&
			\sum_{p=0}^{s-(s_1-s_2)-2}
		a_p
		\Bigg[ \left(	\partial^p
		\mathcal{D}^+
		\mathcal{W}_{s_1} \right)_{\alpha(s_1-s_2+p)\beta(s_1+s_2-1)\dot{\alpha}(p)}
		\left( \partial^{s-p-2} \mathcal{D}^- \mathcal{W}_{s_2} \right)^{\beta(s_1+s_2-1)}_{\alpha(s-s_1+ s_2-p-2)\dot{\alpha}(s- p -2)}
		\\&
		\qquad\qquad\qquad\qquad
		-
		\left(\partial^p
		\mathcal{D}^-
		\mathcal{W}_{s_1}\right)_{\alpha(s_1-s_2+p)\beta(s_1+s_2-1)\dot{\alpha}(p)}
		\left( \partial^{s-p-2}
				\mathcal{D}^+
		\mathcal{W}_{s_2}
		\right)^{\beta(s_1+s_2-1)}_{\alpha(s-s_1+ s_2-p-2)\dot{\alpha}(s- p-2 )}
		\Bigg]
		\\
		&
		+	\sum_{p=0}^{s-(s_1-s_2)-2}
		b_p
		\left(\partial^p
		\mathcal{D}^+	\mathcal{D}^-
		\mathcal{W}_{s_1}
		\right)_{\alpha(s_1-s_2+p+1)\beta(s_1+s_2-1)\dot{\alpha}(p)}
		\left( 
		\partial^{s-p-2}
		\mathcal{W}_{s_2}
		\right)^{\beta(s_1+s_2-1)}_{\alpha(s-s_1+s_2-p-3)\dot{\alpha}(s- p-2)}
			\\
		&
		+	\sum_{p=0}^{s-(s_1-s_2)-2}
		c_p
		\left(\partial^p
		\mathcal{W}_{s_1}
		\right)_{\alpha(s_1-s_2+p-1)\beta(s_1+s_2-1)\dot{\alpha}(p)}
		\left( 
		\partial^{s-p-2}
		\mathcal{D}^+	\mathcal{D}^-
		\mathcal{W}_{s_2}
		\right)^{\beta(s_1+s_2-1)}_{\alpha(s-s_1+s_2-p-1)\dot{\alpha}(s- p-2)}
		.
	\end{split}
}
\end{equation}	
As in the $\mathcal{N}=1$ case we require $|s_1-s_2| \in \mathbb{Z}$.
This ansatz is, by construction, harmonically independent $\mathcal{D}^{++} \mathscr{J}_{\alpha(s-2)\dot{\alpha}(s-2)} = 0 $ and exists only if $s \geq s_1 - s_2 + 2$.

\smallskip

For compactness of the subsequent formulas, we introduce the notation $\Delta = s_1-s_2$.  Then the conservation equations take the form:
\begin{subequations}\label{eq: N=2 conservation}
\begin{equation}
	\mathcal{D}^{+\alpha} \mathscr{J}_{\alpha(s-2)\dot{\alpha}(s-2)} 
	\approx 0
	\qquad
	\Rightarrow
	\qquad
	\begin{cases}
	(-1)^{2s_1+1} a_p \cdot \frac{\Delta+p}{s-2}
	- c_p 
	\cdot 
	\frac{s-\Delta-p-1}{s-2}
	= 0,
	\\
	a_p \cdot \frac{s-\Delta-p-2}{s-2}
	 +
	  (-1)^{2s_1} b_p
	\cdot
	 \frac{\Delta+p+1}{s-2}
	 = 0.
	\end{cases}
\end{equation}	
\begin{equation}
	\bar{\mathcal{D}}^{+\dot{\alpha}}
	\mathscr{J}_{\alpha(s-2)\dot{\alpha}(s-2)} 
	\approx 0
		\qquad
	\Rightarrow
	\qquad
	\begin{cases}
		(-1)^{2s_1+1} a_p
		\cdot
		 \frac{p}{s-2}
		+ 
		b_{p-1} 
			\cdot
		\frac{s-p-3}{s-2}
		= 0,
		\\
		a_{p-1} 
		\cdot
		\frac{s-p-3}{s-2}
		+
		 (-1)^{2s_1+1} c_p 
		 \cdot \frac{p}{s-2}
		 = 0.
	\end{cases}
\end{equation}	
\end{subequations}
As the unique solution of the recurrence system of equations \eqref{eq: N=2 conservation} we obtain
\begin{subequations}
\begin{equation}
	a_p = c \cdot (-1)^p \cdot  \binom{s-2}{\Delta + p}
	\binom{s-4}{p},
\end{equation}	
\begin{equation}
	b_p = c \cdot (-1)^{2s_1-1}
	\cdot (-1)^p 
	 \cdot 
\binom{s-2}{\Delta + p+1}
\binom{s-4}{p},
\end{equation}	
\begin{equation}
	c_p = c \cdot (-1)^{2s_1+1} 
		\cdot (-1)^p 
	\cdot 
	\binom{s-2}{\Delta + p - 1}
	\binom{s-4}{p}.
\end{equation}	
\end{subequations}
Note that the factor \(\binom{s-4}{p}\) in the above solution reflects the
structure of the recurrence relations rather than an additional restriction
on the superspins. For \(s\ge4\) it truncates the sum at \(p\le s-4\), so
that for \(s=4\) only the \(p=0\) terms survive. For \(s=3\) the coefficient
is understood in the generalized sense, \(\binom{-1}{p}=(-1)^p\), which is
non-zero for all \(p\); the sum is then cut off by the upper limit
\(p\le 1-\Delta\). Since \(s\ge s_1-s_2+2\) requires \(\Delta=0,1\) for
\(s=3\), only two cases occur. For \(\Delta=1\) only the \(p=0\) term
survives,
\begin{equation}
	s=3,\ \Delta=1:\qquad
	a_0=c,\qquad b_0=0,\qquad c_0=(-1)^{2s_1+1}c,
\end{equation}
while for \(\Delta=0\) both \(p=0\) and \(p=1\) contribute,
\begin{equation}
	s=3,\ \Delta=0:\qquad
	\begin{aligned}
		&a_0=c,\qquad a_1=c,\\
		&b_0=(-1)^{2s_1+1}c,\qquad b_1=0,\\
		&c_0=0,\qquad c_1=(-1)^{2s_1+1}c.
	\end{aligned}
\end{equation}

\smallskip

In the case of identical superfields $\mathbf{s_1} = \mathbf{s_2}$, unlike the $\mathcal{N}=0,1$ cases analyzed above, the constructed supercurrent does not vanish identically. This can be illustrated, for instance, in the case of integer spins. In the bosonic sector, the $(\mathbf{s}, \mathbf{s_1}, \mathbf{s_2})$ vertex contains
 \begin{equation}
	\begin{split}
		(\mathbf{s}, \mathbf{s_1}, \mathbf{s_2})
		\quad
		\to
		\quad
		&
		(s-1, s_1, s_2),
		\quad
		(s, s_1, s_2-1),
		\quad
		(s, s_1-1, s_2).
	\end{split}
\end{equation} 
From this we see that if one of the bosonic vertices vanishes, then two other non-trivial vertices remain. This is directly related to the fact that 
$\mathcal{N}=2$ supermultiplets combine both integer and half-integer 
$\mathcal{N}=1$ superspins and the three bosonic vertices probe different 
$\mathcal{N}=1$ sectors and cannot all vanish simultaneously.

\smallskip

In the case where the inequality
$$k = s_1 +s_2 - s -1 \geq 0$$
 holds, the vertex can be represented in a manifestly gauge-invariant form and is given by the following chiral action
\begin{equation}\label{eq: N=2 weyls act}
	S^{\text{chiral}}_{(\mathbf{s}, \mathbf{s_1}, \mathbf{s_2})} = \int d^4x d^4\theta \; 	\mathcal{W}^{\alpha(2s-2)}
	\mathcal{W}_{\alpha(2s_1-2-k)\beta(k)}
	\mathcal{W}_{\alpha(2s_2-2-k)}^{\beta(k)}.
\end{equation}	

As in the previous sections, the real and imaginary parts of the constructed complex $\mathcal{N}=2$ principal supercurrent correspond to parity-even and parity-odd vertices, respectively.

\section{Conclusion}\label{sec: concl}

In this paper we have constructed, in manifest Lorentz-covariant form, the
conserved higher-spin currents that generate Born--Infeld-type
\((s,s_1,s_2)\) vertices with the maximal number of derivatives
\(s+s_1+s_2\), for arbitrary \(s\ge s_1\ge s_2\), as well as their
\(\mathcal{N}=1\) and \(\mathcal{N}=2\) supersymmetric extensions. These
vertices are known to be necessary for constructing consistent
interactions at higher orders \cite{Metsaev:1991mt}, and their explicit
form has been missing so far. The central observation behind our
construction is that Born--Infeld vertices are generated by conserved
currents built from Weyl tensors of the same chirality,
\(J\sim \partial C\partial C\), in contrast to the
Berends--Burgers--van Dam currents \(J\sim \partial C\partial\bar{C}\).
Such same-chirality (super)currents were unstudied before; the
results below provide their systematic description at the maximal
number of derivatives.

The interaction is encoded in a conserved
higher-spin current and its trace. We derived the recurrence relations
for the coefficients of the general ansatz, showed how the trace part is
fixed by the conservation equation, and demonstrated that after removing
fake vertices only one nontrivial complex conserved traceless current
remains
\begin{equation}\label{eq: mathbf J res}
	\begin{split}
		\mathbf{J}^{(0)}_{\alpha(s)\dot{\alpha}(s)}
		=
		\sum_{p=0}^{s- (s_1-s_2)}
		(-1)^p
		\frac{\binom{s}{s_1-s_2+p} \binom{s}{p}}{\binom{s}{s_1-s_2}} &\left(\partial^p C_{s_1}\right)_{\alpha(s_1-s_2+p) \beta(s_1+s_2) \dot{\alpha}(p)}
		\\
		&\times
		\left( \partial^{s-p} C_{s_2} \right)^{\beta(s_1+s_2)}_{\alpha(s+s_2-s_1-p)\dot{\alpha}(s-p)}.
	\end{split}
\end{equation}	
 Its real and imaginary parts correspond
to parity-even and parity-odd interactions, respectively. 
For the case of identical spins $s_1$ and $s_2$, for odd $s$ the constructed current vanishes identically.  
In the case
\(s\le s_1+s_2\), the same vertex was rewritten in a manifestly
gauge-invariant form in terms of higher-spin Weyl tensors, thereby
establishing the equivalence of the current and Weyl-tensor
representations.

We then generalized the construction to supersymmetric theories. For
\(\mathcal{N}=1\) supersymmetry, we constructed the general conserved 
higher-spin supercurrent of integer superspin
 $s$ with manifest supersymmetry and found the
corresponding \(\mathcal{N}=1\) Born--Infeld-type cubic vertices. The resulting supercurrent reads
  \begin{equation}
	{\scriptsize 	\begin{split}
	\mathcal{J}_{\alpha(s)\dot{\alpha}(s-1)}
		=
		&
		\sum^{s-(Y_1-Y_2)-1}_{p=0} 
		(-1)^p \cdot \binom{s-1}{p} \binom{s}{\Delta+p+1}		
		\left(\mathcal{D} \partial^p \mathcal{W}_{Y_1}\right)_{\alpha(Y_1-Y_2+p+1)\beta(Y_1+Y_2)\dot{\alpha}(p)} 
		\\
		&
		\qquad\qquad\qquad\qquad\qquad\qquad\qquad\qquad\qquad
		\times
		\left(
		\partial^{s-p-1} \mathcal{W}_{Y_2}
		\right)^{\beta(Y_1+Y_2)}_{\alpha(s-(Y_1-Y_2)-p-1) \dot{\alpha}(s-p-1)}
		\\
		&
		-
		\sum_{p=0}^{s-(Y_1-Y_2)-1} 
		(-1)^{2Y_1  } \cdot (-1)^p \cdot \binom{s-1}{p} \binom{s}{\Delta+p}
		\left(\partial^p \mathcal{W}_{Y_1}\right)_{\alpha(Y_1-Y_2+p)\beta(Y_1+Y_2)\dot{\alpha}(p)} 
		\\
	&
	\qquad\qquad\qquad\qquad\qquad\qquad\qquad\qquad\qquad
	\times
		\left(
		\mathcal{D} 
		\partial^{s-p-1} \mathcal{W}_{Y_2}
		\right)^{\beta(Y_1+Y_2)}_{\alpha(s-(Y_1-Y_2)-p) \dot{\alpha}(s-p-1)}.
	\end{split}}
\end{equation}	
In the case of identical superspins $Y_1= Y_2$ this reproduces the result of \cite{Gates:2019cnl}, while for arbitrary $Y_1, Y_2$
it provides the explicit form of the supercurrent, which was not previously available.
The supercurrent vanishes identically for
odd integer superspin \(s\), the supersymmetric counterpart of the vanishing of the bosonic current for odd $s$. For \(s\le Y_1+Y_2\), the interaction
admits a chiral superspace representation in terms of \(\mathcal{N}=1\)
higher-spin Weyl supertensors. 

For \(\mathcal{N}=2\) supersymmetry, using
the harmonic-superspace formulation, we constructed the complex principal
higher-spin supercurrent and showed that the conservation equations fix its
coefficients uniquely. The principal supercurrent takes the form:
 \begin{equation}
	{\footnotesize
		\begin{split}
			&	\mathscr{J}_{\alpha(s-2)\dot{\alpha}(s-2)} 
			=
			\\
			&
			\sum_{p=0}^{s-(s_1-s_2)-2}
			(-1)^p \cdot  \binom{s-2}{\Delta + p}
			\binom{s-4}{p}
			\\
			&\qquad\qquad\quad\quad
			\times
			\Bigg[ \left(	\partial^p
			\mathcal{D}^+
			\mathcal{W}_{s_1} \right)_{\alpha(s_1-s_2+p)\beta(s_1+s_2-1)\dot{\alpha}(p)}
			\left( \partial^{s-p-2} \mathcal{D}^- \mathcal{W}_{s_2} \right)^{\beta(s_1+s_2-1)}_{\alpha(s-s_1+ s_2-p-2)\dot{\alpha}(s- p -2)}
			\\&
			\qquad\qquad\qquad\qquad
			-
			\left(\partial^p
			\mathcal{D}^-
			\mathcal{W}_{s_1}\right)_{\alpha(s_1-s_2+p)\beta(s_1+s_2-1)\dot{\alpha}(p)}
			\left( \partial^{s-p-2}
			\mathcal{D}^+
			\mathcal{W}_{s_2}
			\right)^{\beta(s_1+s_2-1)}_{\alpha(s-s_1+ s_2-p-2)\dot{\alpha}(s- p-2 )}
			\Bigg]
			\\
			&
			+	
			(-1)^{2s_1-1}
			\sum_{p=0}^{s-(s_1-s_2)-2}
			(-1)^p 
			\binom{s-2}{\Delta + p+1}
			\binom{s-4}{p} 
			\left(\partial^p
			\mathcal{D}^+	\mathcal{D}^-
			\mathcal{W}_{s_1}
			\right)_{\alpha(s_1-s_2+p+1)\beta(s_1+s_2-1)\dot{\alpha}(p)}
			\\
			&\qquad\qquad\qquad\qquad
			\qquad\qquad\qquad\qquad
			\qquad\qquad\qquad\qquad
			\times
			\left( 
			\partial^{s-p-2}
			\mathcal{W}_{s_2}
			\right)^{\beta(s_1+s_2-1)}_{\alpha(s-s_1+s_2-p-3)\dot{\alpha}(s- p-2)}
			\\
			&
			+	
			(-1)^{2s_1+1} 
			\sum_{p=0}^{s-(s_1-s_2)-2}
			(-1)^p 
			\binom{s-2}{\Delta + p - 1}
			\binom{s-4}{p}
			\left(\partial^p
			\mathcal{W}_{s_1}
			\right)_{\alpha(s_1-s_2+p-1)\beta(s_1+s_2-1)\dot{\alpha}(p)}
			\\
			&\qquad\qquad\qquad\qquad
			\qquad\qquad\qquad\qquad
			\qquad\qquad\qquad\qquad
			\times
			\left( 
			\partial^{s-p-2}
			\mathcal{D}^+	\mathcal{D}^-
			\mathcal{W}_{s_2}
			\right)^{\beta(s_1+s_2-1)}_{\alpha(s-s_1+s_2-p-1)\dot{\alpha}(s- p-2)}
			.
		\end{split}
	}
\end{equation}	
Unlike the \(\mathcal{N}=0\) and \(\mathcal{N}=1\)
cases with identical fields, the \(\mathcal{N}=2\) vertex with identical
superfields does not vanish identically; this is consistent with the
component decomposition, in which several bosonic vertices contribute.

\smallskip

Taken together, these results give a uniform description of
Born--Infeld vertices in the non-supersymmetric, \(\mathcal{N}=1\) and
\(\mathcal{N}=2\) cases.
The explicit form of the currents obtained here opens the way to several
further developments. Since Born--Infeld vertices are known to be
necessary for constructing consistent interactions at higher orders
\cite{Metsaev:1991mt}, the currents above provide a convenient starting
point for analyzing higher-order deformations of the cubic action.
It would also be interesting to extend the construction to anti-de Sitter
background and to \(\mathcal{N}\ge3\) supersymmetry, where the relevant
superfield formalisms are less developed.

\acknowledgments

The author is grateful to Yu.~Zinoviev for discussions of cubic vertices
and E.~Ivanov for many fruitful discussions and support. 
This work was partially supported by
the Foundation for the Advancement of Theoretical Physics and
Mathematics ``BASIS'', grant \verb|#| 25-1-1-10-4 .

\appendix


\end{document}